\documentclass{article}
\usepackage{graphicx}
\usepackage{amsmath}
\usepackage{appendix}
\usepackage{multicol}
\usepackage{fullpage}
\usepackage{hyperref}
\hypersetup{
	colorlinks   = true, %Colours links instead of ugly boxes
	urlcolor     = blue, %Colour for external hyperlinks
	linkcolor    = blue, %Colour of internal links
	citecolor    = blue  %Colour of citations
}
\usepackage{epigraph}
\usepackage{setspace}
\begin{document}

	\title{\huge{ \textbf{Adjoint Method for Macroscopic Phase-Resetting Curves of Generic Spiking Neural Networks}}}
	
	\author{Gr\'{e}gory Dumont$^{1}$, Alberto P\'{e}rez-Cervera$^{2}$ and Boris Gutkin$^{1,2}$\\
		\parbox{13.5cm}{
			\small
			\begin{itemize}
				\item[$^1$]
				Group for Neural Theory, LNC INSERM U960, DEC, Ecole Normale Superieure PSL* University, Paris France 
				\item[$^2$]
				Center for Cognition and Decision Making, Institute for Cognitive
				Neuroscience, National Research University Higher School of Economics,
				Moscow
			\end{itemize}
	}}

	\date{}
	\maketitle
	
	\noindent \textbf{Corresponding author:} Gr\'{e}gory Dumont, \texttt{gregory.dumont@ens.fr} \\

	\begin{abstract}
		Brain rhythms emerge as a result of synchronization among interconnected spiking neurons. Key properties of such rhythms can be gleaned from the phase-resetting curve (PRC). Inferring the macroscopic PRC and developing a systematic phase reduction theory for emerging rhythms remains an outstanding theoretical challenge. Here we present a practical theoretical framework to compute the PRC of generic spiking networks with emergent collective oscillations. To do so, we adopt a refractory density approach where neurons are described by the time since their last action potential. In the thermodynamic limit, the network dynamics are captured by a continuity equation known as the refractory density equation. We develop an appropriate adjoint method for this equation which in turn gives a semi-analytical expression of the infinitesimal PRC. We confirm the validity of our framework for specific examples of neural networks.  Our theoretical findings highlight the relationship between key biological properties at the individual neuron scale and the macroscopic oscillatory properties assessed by the PRC.
	\end{abstract}
	
	\vspace{1.5cm}
	
	Popularized by Arthur T. Winfree in 1980 \cite{winfree}, the phase-resetting curve (PRC) has been one of the central tools to study properties and mechanisms of biological rhythms.  The PRC is a measure that tracks down the phase shift of an oscillation when a transient perturbation is presented at a determined phase of the oscillatory cycle. It is particularly well adapted to clarify essential dynamical features of measured data in a wide variety of biological contexts, see for instance \cite{Stiefeljn2015} for data in neuroscience. The multiple advantages of using the PRC  have been summarized in multiple works \cite{ Ashwin2016, Nakao2016}. For instance, it has proven to be especially efficient to predict the phase-locking behavior of coupled neural oscillators \cite{Achuthan5218}, and to study information flow in networks of bio-chemical oscillators \cite{Kirst2016}.

	For oscillations that can be expressed as ordinary differential equations, the adjoint method, see \cite{Brown2004}, provides an accurate procedure to compute the so-called infinitesimal PRC (iPRC). In the case of vanishingly small perturbation amplitudes, PRC and iPRC become proportional to each other, and therefore, for a perturbation that is small enough, any oscillating dynamical system can be reduced to a single phase equation:
	\begin{equation*}
		\begin{split}
			\dfrac{d}{dt}\theta(t)=\omega +Z(\theta(t)) \cdot p(t).
		\end{split}
	\end{equation*}
	Here $\theta$ is the oscillation phase, $\omega$ is the natural frequency of the oscillator, $p(t)$ represents the time dependent-perturbation, and the function $Z$ the iPRC computed via the adjoint method.

	There are multiple reasons to use the PRC to characterize brain oscillations and it has been the subject of recent discussion   \cite{Canavier2015} and experimental setups \cite{Akam2018}.
	However, in the brain, most rhythms emerge from the interaction of irregular spiking cells \cite{buzsaki2006rhythms}. Hence the brain oscillatory activity is a consequence of synchronisation among firing events of a large population of neurons that can not be portrayed by elementary dynamical systems. Although first steps toward deriving macroscopic PRCs for emergent oscillations have been made, e.g. \cite{DumontEG2017,Akao2018}, these efforts require rather restrictive assumptions on the network models considered and extracting the iPRC of generic oscillating spiking networks has remain elusive so far. In this letter, we address the need to go beyond the traditional adjoint method and move toward a framework that permits the iPRC computation of generic spiking circuits.

	Our approach relies on a mean-field description of networks where a given cell is characterized by the amount of time passed by since its last action potential. There are undoubtedly alternative ways to describe neurons, however, such a formalism is general as it can effectively  reflect many spiking formulations. For instance, renewal processes such as the noisy integrate-and-fire \cite{dumont2016noisy,dumont2016theoretical}, or spike response models \cite{Gerstner2000}, can be expressed within this framework. Furthermore, this approach provides approximation schemes for complex biophysically-realistic models \cite{Chizhov2007,Chizhov2014}, for correlated noise \cite{Chizhov2008}, and for neural adaptation \cite{Naud2012,Deger2014}, see \cite{SCHWALGER2019155} for a recent review.

	In the thermodynamic limit, the network is well represented by a partial differential equation known as the von Foerster equation \cite{VonF1959}; a continuity equation for which several textbooks in mathematical biology devote an entry \cite{murray2002,MB,perthame}. In the neuroscience community, we refer to this continuity equation as the refractory density equation. It was first implemented by Wulfram Gerstner and  Leo van Hemmen in 1992 \cite{Gerstner1992}. The refractory equation can rigorously be derived starting from the stochastic process \cite{Chevalier2015}, and is amenable to mathematical analysis \cite{Pakdaman2010}.  Moreover, this continuity equation has been a major tool for studying emergent synchronized assemblies \cite{Gerstner2000}, transient dynamics \cite{Deger2010}, low dimensional reduction \cite{Pietras2020}, and finite-size network activity fluctuations \cite{Meyer2002,Deger2014,Schwalger2017,dumont2017}. We recommend the reader the textbook \cite{gerstner2014neuronal} for an intuitive introduction on the refractory density equation.
	
	Here, we develop an adjoint method for the refractory density equation and compare the PRC of the full network to the analytically obtained iPRC. We illustrate our theoretical finding using a typical scenario with oscillations emerging from a recurrent excitatory neural network. We also discuss the generalization of our results to more complex network architectures, such as an excitatory-inhibitory network, in the supplementary information.

	Let us start with spiking neurons  that are described as renewal processes. It takes into account $h(t)$, the total input a neuron receives  and $r$, the time since the last action potential. Denoting $S(h(t),r)$ the escape rate, then, the probability that a firing event occurs during a time internal  $dt$ is given by $S(h(t),r)dt$. 
	Note that the escape rate reflects the individual properties of  neurons, as an example, we take an escape rate that captures the dynamics of pyramidal cells  \cite{gerstner2014neuronal}.
	As soon as an action potential is triggered, the neuron's age $r$ is reset to zero.
	The population activity can be extracted and is given by the sum of all the occurring spikes:

	\begin{figure}
		\begin{center}  
			\includegraphics[width=0.9\linewidth]{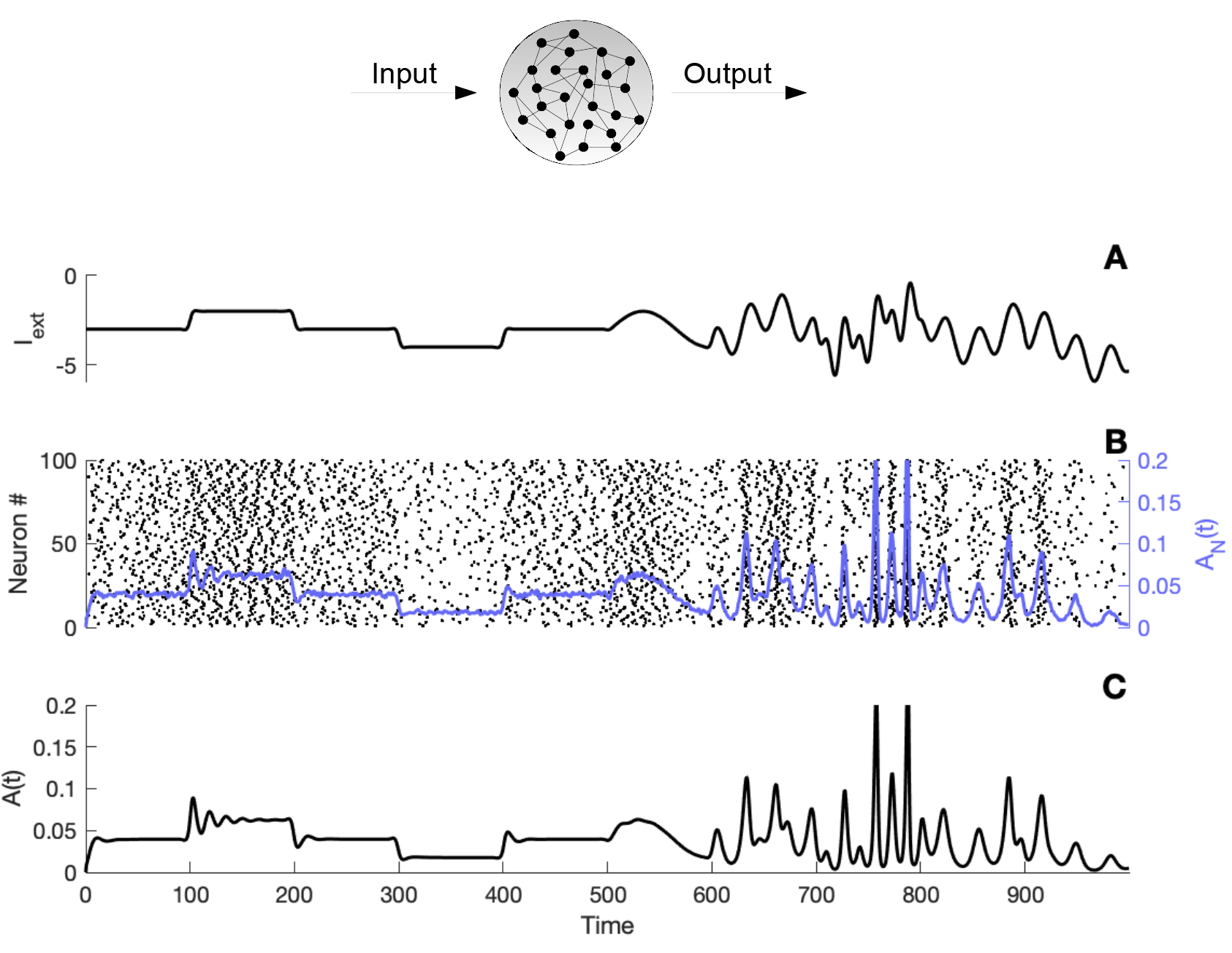}
			\caption{Dynamics for a recurrent excitatory network. Top panel: Schematic illustration of the network. The network receives an input, $I_{ext}(t)$ and produces a firing activity $A(t)$, considered to be the output of the network. Lower panels: Comparison of firing activity. A) Time evolution of the stimulus $I_{ext}(t)$. B) Raster plot of $100$ neurons, the blue line displays the resulting firing activity Eq. (\ref{FR_N}) of the full network. C) Firing activity obtained from a simulation of the mean-field equation (\ref{FR}). The simulation was initiated with a similar Gaussian profile for the full network and the mean-field equation, parameters: $S(h,r)=\exp(h)H(r-T_{ref})\left(1-\exp \left(-\left(r-T_{ref}\right) /\tau \right) \right)$, $T_{ref}=10$, $\tau_s=10$, $\tau=5$, $J_s=15$, $N = 5000$ and $\Delta t = 0.05$.   }
			\label{Fig1}
		\end{center}
	\end{figure}

	\begin{equation}
		\label{FR_N}
		A_N(t)= \dfrac{1}{N}  \sum_{k=1}^{N} \sum_{f} \delta(t-t_k^f) .
	\end{equation}
	where $\delta$ is the Dirac mass, $N$ the number of neurons and $t_k^f$ the firing time of the cell numbered $k$.
	The total input current is given by 
	\begin{equation*}
		\begin{split}
			h(t)= I_{ext}(t)   + I_s(t) ,
		\end{split}
	\end{equation*}
	where $I_{ext}(t)$ is an external current and the synaptic $I_s(t)$, which defines the current feedback of the network, is given by
	\begin{equation*}
		\begin{split}
			I_s(t)=   J_s \kappa * A_N(t) .
		\end{split}
	\end{equation*}
	Here $J_s$ is the synaptic efficiency and $\kappa$ the normalized synaptic filter
	\begin{equation*}
		\begin{split}
			\kappa (t)=\dfrac{e^{-t/ \tau_s}}{\tau_s},
		\end{split}
	\end{equation*}
	with $\tau_s$ the synaptic decay.

	In the limit of an infinitely large number of neurons $N$ (the thermodynamic limit), the full network description reduces to a single partial differential equation. 
	Denoting $q(t,r)$ the probability density for a neuron to have at time $t$ an age $r$, the density profile evolves according to the continuity equation:
	\begin{equation}\label{AS}
		\begin{split}
			\frac{\partial}{\partial t}q(t,r) +\frac{\partial}{\partial r}q(t,r)=-S(h(t),r)q(t,r).
		\end{split}
	\end{equation}
	Because once a cell emits an action potential its age is reset to zero, the
	natural boundary condition is
	\begin{equation*}
		q(t,0)=A(t),
	\end{equation*}
	where $A(t)$ is the neural network activity and is defined as
	\begin{equation}
		\label{FR}
		A(t)= \int_{0}^{+\infty}S(h(t),r)q(t,r) \,dr .
	\end{equation}
	The total input current is still given by
	\begin{equation*}
		\begin{split}
			h(t)= I_{ext}(t)   + I_s(t) ,
		\end{split}
	\end{equation*}
	but this time the synaptic current $I_s(t)$ is computed as
	\begin{equation*}
		\begin{split}
			I_s(t)=   J_s \kappa * A(t) .
		\end{split}
	\end{equation*}
	
	The mean-field equation (\ref{AS})  above defines a conservation law and expresses three different processes taking place at the cellular level: a drift process due to the time passing between action potentials,
	an escape rate generated by the randomness of firing events and the individual cell properties, a non-local boundary condition which describes the reset of the neurons that just fired.

	The numerical simulations presented in Fig.~\ref{Fig1} illustrate a comparison between the dynamics of the full network activity and the mean-field equation (\ref{FR}).  The figure shows that the mean-field  captures the essential shape of the full network firing activity.

	\begin{figure}
		\begin{center}
			\includegraphics[width=0.85\linewidth]{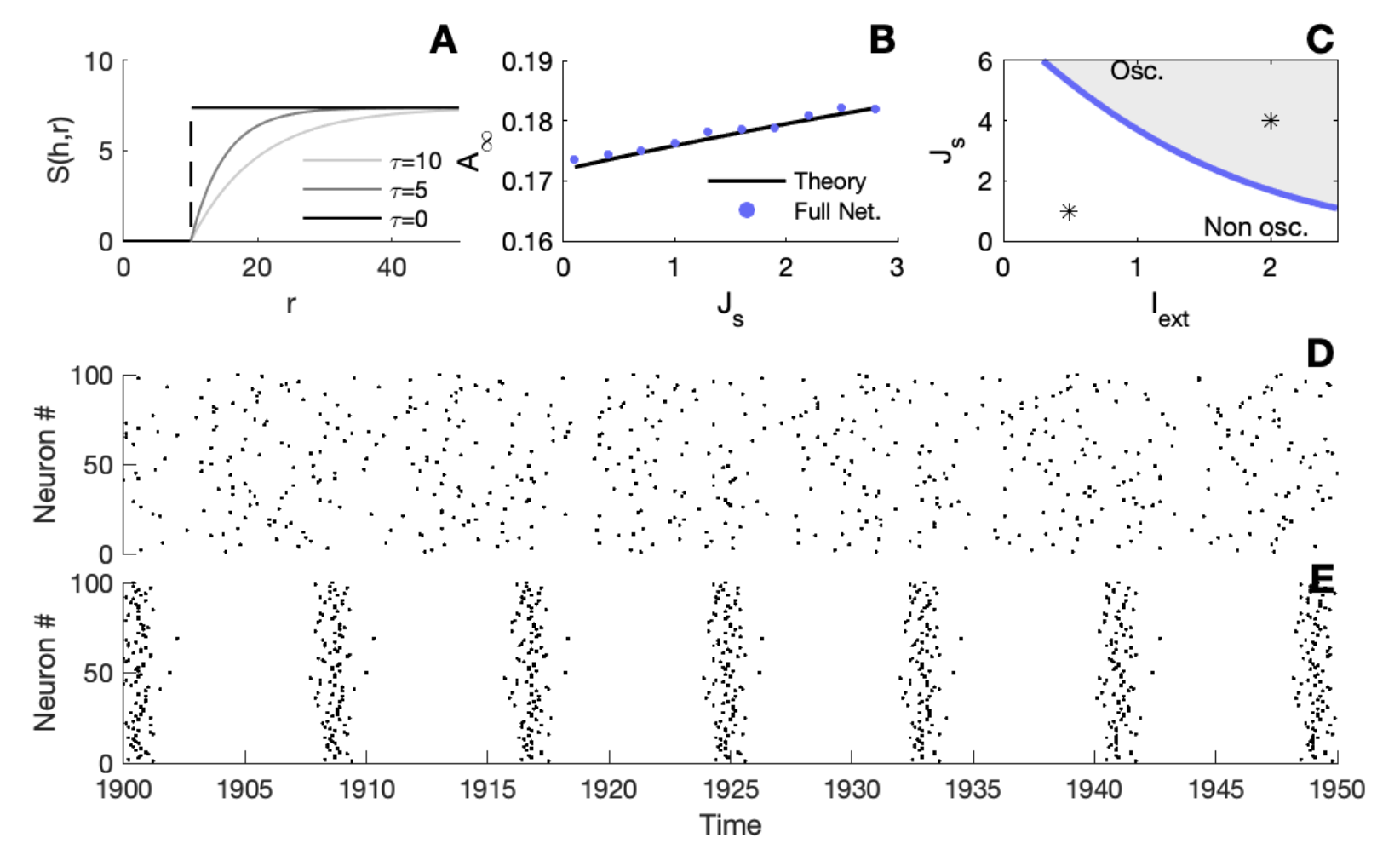}
			\caption{Emergent oscillations. A)  Illustration of the escape rate $S(h,r)$ for different value of the parameter $\tau$, ($h=2$). B) Comparison between the steady state firing activities, blue dots for the full network, and the black line for the theoretical prediction given by (\ref{MFT}). C) Bifurcation line in the parameter space (blue curve). The grey shaded region corresponds to an oscillatory regime of the neural network, the white region corresponds to a stable asynchronous mode of the network. D) and E) Raster plots of the spiking activity of $100$ neurons. Panel B corresponds to the black asterisk lying in the asynchronous (white) region of panel A, whereas panel C depicts the activity that corresponds to the black asterisk lying in the oscillatory (grey) region of panel A. parameters: $S(h,r)=\exp(h)H(r-T_{ref})\left(1-\exp \left(-\left(r-T_{ref}\right) /\tau \right) \right)$, $T_{ref}=10$, $\tau_s=10$, $\tau=0$, $N = 5000$ and $\Delta t = 0.1$.  }
			\label{Fig2}
		\end{center}
	\end{figure}

	To investigate the emergence of macroscopic oscillations we perform a nonlinear analysis of the mean-field density equation (\ref{AS}). 
	After algebraic manipulations - see supplementary information (SI) for details - we find that the steady state is given by
	\begin{equation*}
		\begin{split}
			q_{\infty}(r)= A_{\infty} e^{- \int_0^r  S_{\infty}(s) \,ds  },
		\end{split}
	\end{equation*}
	where the mean activity in the asynchronous regime $A_\infty$ and the mean input $h_{\infty}$ are given by
	\begin{equation}\label{MFT}
		\begin{split}
			A_{\infty}^{-1}= \int_0^{+\infty} e^{- \int_0^r  S_{\infty}(s) \,ds  }  \,dr ,\quad h_{\infty} = I_{ext} +J_s A_{\infty},
		\end{split}
	\end{equation}
	note that we have used the notation
	\begin{equation*}
		\begin{split}
			S_{\infty}(r) :=S(h_{\infty},r).
		\end{split}
	\end{equation*}

	Linearizing around the steady state we can extract the characteristic equation, whose solutions give the eigenvalues. The time-independent solution will loose its stability as soon as there is an eigenvalue having a positive real part. The characteristic equation reads
	\begin{equation*}
		\begin{aligned}
			\mathcal{C} (\lambda) =& \enskip J_s  \hat{\kappa}_\lambda   \int_{0}^{+\infty}S_{\infty} \int_0^{r}  \frac{\partial S_{\infty}}{\partial h} q_{\infty} e^{- \int_x^{r}  S_{\infty}   +\lambda  \,ds } \,dx  \,dr \\
			&+ 1-J_s  \hat{\kappa}_\lambda  \int_{0}^{+\infty}  \frac{\partial S_{\infty}}{\partial h} q_{\infty} \,dr  
			-  \int_{0}^{+\infty}S_{\infty}  e^{- \int_0^{r}  S_{\infty}  +\lambda  \,ds } \,dr  
		\end{aligned}
	\end{equation*}
	where $\ \hat{\kappa}_\lambda$ is the Laplace transform of the synaptic filter $\kappa$, see SI for details of the computations.

	The bifurcation line, which separates an oscillatory dynamic from an asynchronous steady-state regime, can be obtained numerically from the characteristic equation by solving:
	\begin{equation*}
		\mathcal{C} (i \omega) = 0.
	\end{equation*}
	As we can see from Fig.~\ref{Fig2}C, for sufficiently large synaptic strengths $J_s$ and the external current $I_{ext}$, the asynchronous state undergoes a bifurcation. The simulated spiking activity of the full network in Fig.~\ref{Fig2}D-E confirms the emergence of a transition from an asynchronous to a synchronized activity regime when parameters are taken in the respective side of the bifurcation line.

\begin{figure}
		\begin{center}
			\includegraphics[width=0.85\linewidth]{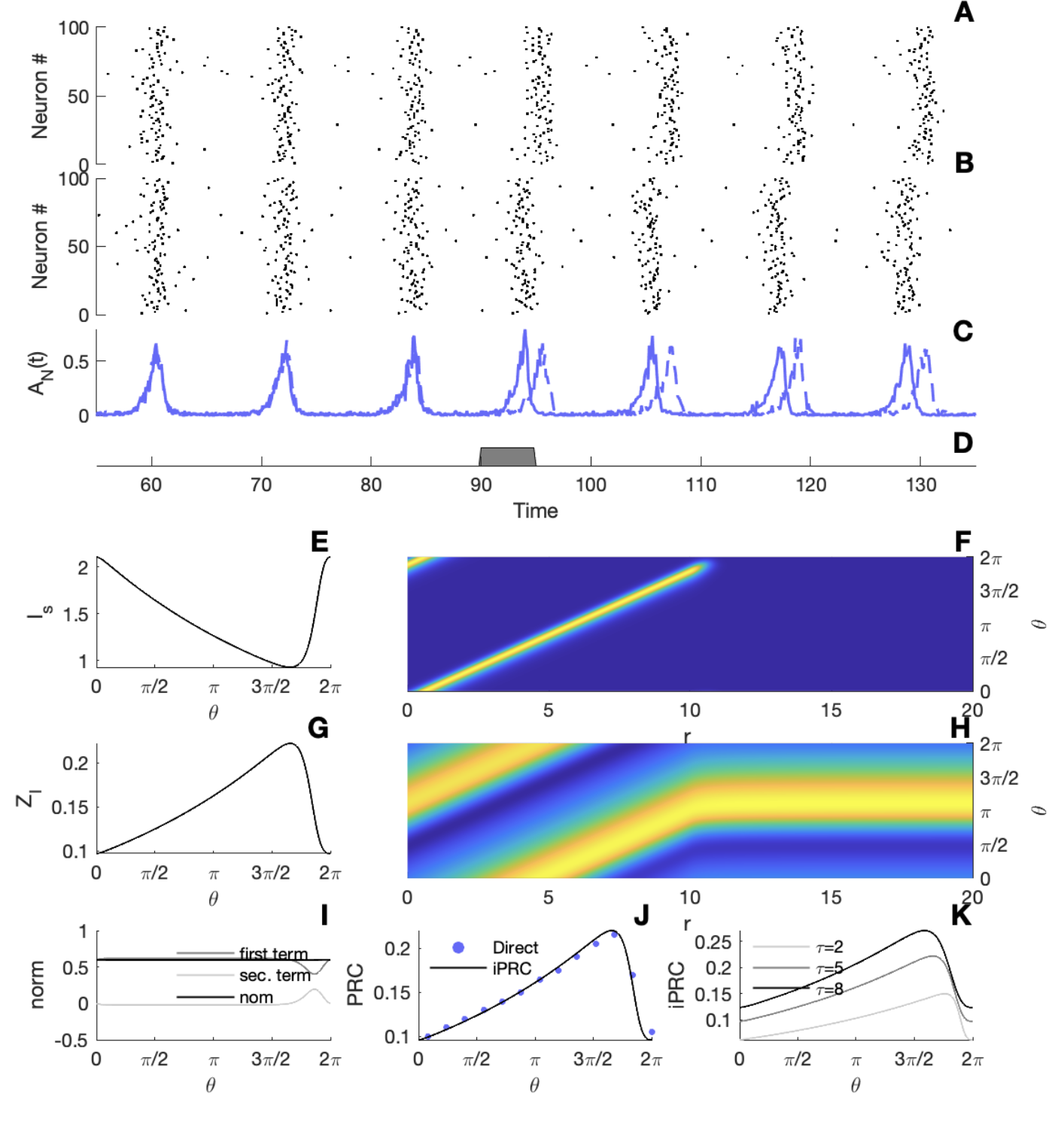}
			\caption{Phase-resetting curve.  A-B) Raster plot of $100$ neurons from a simulation  of  a non-perturbed/perturbed  network. C) Resulting firing activity of the networks obtained from Eq. (\ref{FR_N}), the dashed line for the non-perturbed network and full line for the perturbed one. D)  Illustration of the stimulus. E-F) The panels give the periodic solution of the mean-field equation (\ref{AS}). G-H) The panels give the periodic solution of the adjoint system (\ref{iPRC1})-(\ref{iPRC2}). I) Illustration of normalizing condition (\ref{normalisation}). 
				J) The panel displays the macroscopic PRC, the black line illustrates the solution of Eq.~(\ref{iPRC1}), while blue dots indicate the PRC obtained via direct perturbations. K) Solution of the iPRC (\ref{iPRC2}) for different values of the parameter $\tau$. Parameters: $S(h,r)=\exp(h)H(r-T_{ref})\left(1-\exp \left(-\left(r-T_{ref}\right) /\tau \right) \right)$, $I_{ext}=2$, $T_{ref}=10$, $\tau_s=10$, $\tau=5$, $J_s=15$, $N = 5000$ and $\Delta t = 0.005$.  Direct perturbations in panel D were made with a square wave current pulse (amplitude $3$, duration $5$) on the full network, and in panel J with a square wave (amplitude $8$, duration $0.8$) on the mean-field system (\ref{AS}).}
			\label{Fig3}
		\end{center}
	\end{figure}

	When a brief depolarizing current is applied to the network in the oscillatory regime,  the firing activity will shift  in  time. Raster plots from numerical simulations of the full network illustrate the macroscopic phase shift (Fig. \ref{Fig3}A-B); notice the resulting phase shift in the firing activity displayed in Fig. \ref{Fig3}C.

	We are now ready to construct the iPRC, defined  mathematically  for  an infinitesimally small perturbation. We find, see SI for details, the iPRC to be solution of an associated adjoint mean-field equation
	\begin{equation}\label{iPRC1}
		-\frac{\partial }{\partial t} Z_q (t,r)  - \frac{\partial}{\partial r}Z_q(t,r) 
		= - S(h_{o}(t),r)  \left[ Z_q (t,r)  - Z_q (t,0) -   \dfrac{J_s}{\tau_s} Z_{I_s}(t)  \right] ,
	\end{equation}
	and
	\begin{equation}\label{iPRC2}
		-   \frac{d }{d t}  Z_{I_s} (t)   =-  \dfrac{1}{\tau_s} Z_{I_s}(t)     
		-   \int_{0}^{+\infty}\left[ Z_q (t,r)  - Z_q (t,0) -   \dfrac{J_s}{\tau_s} Z_{I_s}(t)  \right]  \frac{\partial S}{\partial h} (h_{o}(t),r)q_{o}(t,r)   \,dr,
	\end{equation}
	where
	\begin{equation*}
		h_o(t)=I_{ext}+I_{s_o}(t).
	\end{equation*}
	Here, $(q_{o},I_{s_o})$ is the investigated periodic solution of the mean-field equation (\ref{AS}). The iPRC is given by the unique periodic solution, SI, that satisfies the normalizing condition
	\begin{equation}\label{normalisation}
		\int_{0}^{+\infty}  Z_q(t,r) \frac{\partial }{\partial t}q_o(t,r)     \,dr      +  Z_{I_s}(t)  \frac{d }{d t} I_{s_o}(t)    =  \dfrac{2\pi }{T}.
	\end{equation}\\
	When directly applied perturbations are small enough, the PRC and the iPRC become proportional to each other.
	In a real setting, incoming perturbations should get through the synapse, therefore, $Z_{I_s}$ should be interpreted as the  iPRC of the macroscopic oscillation.

	An illustration of the periodic solution (Fig. \ref{Fig3}E-F) and its associated periodic adjoint solution (Fig. \ref{Fig3}G-H) is presented. The adjoint solution has to be normalized according to (\ref{normalisation}) as we can see in Fig. \ref{Fig3}I.  We compare in Fig. \ref{Fig3}J the analytically determined iPRC solution of Eq. (\ref{iPRC2}) to the PRC obtained from direct perturbations of the spiking neural network. It shows an excellent agreement, confirming the validity of our theoretical approach. 
	Note that the PRC depends on cell properties. Indeed, changing  parameters of $S$ alters the shape the iPRC as illustrated in Fig. \ref{Fig3}K . 
	The numerical procedure to solve the adjoint system is presented in SI.
	
	In this letter, we presented a theoretical framework to compute the iPRC of emergent macroscopic network-wide oscillations. We considered generalized spiking networks  that can be used to understand key issues related to emerging brain rhythms with a wide variety of neuronal models. In summary, the methodology presented here can be applied to a wide variety of network models and opens avenues for multiple research direction on the links between the individual component properties (e.g. neural excitability) and collective phenomena. Connections with experimentally measured PRCs is a fruitful future research direction.\\ 
	
	\vspace{1.5cm}
	
	\appendix
	
	\newpage

	%======   APPENDIX A  ==========%
	\textbf{\huge{Supplementary Information}}\\

	Denoting $q(t,r)$ the probability density for a neuron to have at time $t$ an age $r$, the refractory density profile evolves according to the continuity equation:
	\begin{equation}\label{ASsup}
		\begin{split}
			\frac{\partial}{\partial t}q(t,r) +\frac{\partial}{\partial r}q(t,r)=-S(h(t),r)q(t,r).
		\end{split}
	\end{equation}
	The function $S(h(t),r)$ is the escape rate which reflects the individual properties of  neurons. The total input current $h(t)$ is given by 
	\begin{equation*}
		\begin{split}
			h(t)= I_{ext}(t)   + I_s(t) ,
		\end{split}
	\end{equation*}
	where $I_{ext}$ is the external current and $I_s$ the synaptic current: 
	\begin{equation*}
		\begin{split}
			I_s(t)=   J_s \kappa * A(t) .
		\end{split}
	\end{equation*}
	Here $J_s$ is the synaptic efficiency,  $A(t)$ the firing activity defined as
	\begin{equation*}
		A(t)= \int_{0}^{+\infty}S(h(t),r)q(t,r) \,dr ,
	\end{equation*}
	and $\kappa$ the normalized synaptic filter
	\begin{equation*}
		\begin{split}
			\kappa (t)=\dfrac{e^{-t/ \tau_s}}{\tau_s},
		\end{split}
	\end{equation*}
	with $\tau_s$ the synaptic decay.\\
	
	The mean-field equation (\ref{ASsup}) is endowed with a boundary condition:
	\begin{equation*}
		q(t,0)=A(t).
	\end{equation*}
	%

	%======   APPENDIX A  ==========%
	\section{Steady State}\label{A01}

	The asynchronous state can be computed as the time independent solution of the refractory density equation. Let us denote $q_{\infty}(r)$ the steady state, and $A_{\infty} $ the mean firing rate. We have the following equation
	\begin{equation*}
		\begin{split}
			\frac{d}{d r}q_{\infty}(r)=-S(h_{\infty},r)q_{\infty}(r),
		\end{split}
	\end{equation*}
	where we have noted 
	\begin{equation*}
		\begin{split}
			h_{\infty} = I_{ext} + J_s A_{\infty}.
		\end{split}
	\end{equation*}
	The equation can be integrated and gives us
	\begin{equation*}
		\begin{split}
			q_{\infty}(r)= A_{\infty} e^{- \int_0^r  S(h_{\infty},s) \,ds  },
		\end{split}
	\end{equation*}
	where we have used the natural boundary condition
	\begin{equation*}
		\begin{split}
			q_{\infty}(0)= A_{\infty}.
		\end{split}
	\end{equation*}
	Finally, the asynchronous mean firing rate can be computed using the conservation property of the neural network
	\begin{equation*}
		\begin{split}
			\int_0^{\infty} q_{\infty}(r) \,dr =1 ,
		\end{split}
	\end{equation*}
	and we get
	\begin{equation*}
		\begin{split}
			A_{\infty}^{-1}  = \int_0^{\infty} e^{- \int_0^r  S(h_{\infty},s) \,ds  }   \,dr,
		\end{split}
	\end{equation*}
	Note that the mean firing rate is only implicitly given since $h_{\infty}$ does depends on $A_{\infty}$.

	With our choices of functions
	\begin{equation*}
		\begin{split}
			S(h_{\infty},r)  = e^{h_{\infty}} H(r-T_{ref})  ,
		\end{split}
	\end{equation*}
	we can push further the computation, and after algebraic manipulations, we find that the mean firing activity $A_{\infty}$ is solution of the nonlinear equation
	\begin{equation}\label{SSAA}
		\begin{split}
			A_{\infty}   = \left( T_{ref}+ e^{-I_{\infty}  -J_sA_{\infty} }\right)^{-1},
		\end{split}
	\end{equation}
	which can be solved numerically. \\

	%======   APPENDIX B  ==========%
	\section{Stability Analysis}\label{A02}
	To study the stability of the asynchronous state, one needs the eigenvalues of the differential operator once a linearization around the steady state has been performed. We therefore consider a small perturbation and write the solution in the form
	\begin{equation*}
		\begin{aligned}
			q(t,r) = q_{\infty}(r)+ \varepsilon q_1(t,r) +\mathcal{O}( \varepsilon^2 ), \quad A(t) = A_{\infty} + \varepsilon A_1(t) +\mathcal{O}( \varepsilon^2 ). 
		\end{aligned}
	\end{equation*}
	Plugging these expressions into Eq.~(\ref{ASsup}) - keeping the first order terms only - yields the partial differential equation
	\begin{equation*}
		\begin{split}
			\frac{\partial}{\partial t}q_1(t,r) +\frac{\partial}{\partial r}q_1(t,r)=-S(h_{\infty},r)q_1(t,r)  -J_s  \frac{\partial S}{\partial h} (h_{\infty},r)q_{\infty}(r) \kappa * A_1 (t) ,
		\end{split}
	\end{equation*}
	and for the activity
	\begin{equation*}
		\begin{split}
			A_1(t)=\int_{0}^{+\infty}S(h_{\infty},r)q_1(t,r) \,dr +J_s\kappa * A_1 (t)  \int_{0}^{+\infty}  \frac{\partial S}{\partial h} (h_{\infty} , r)q_{\infty}(r) \,dr .
		\end{split}
	\end{equation*}
	Since we are interested in the long term behavior of the perturbation we express the perturbation in eigenvalue mode
	\begin{equation*}
		\begin{split}
			q_1(t,r) =e^{\lambda t}q_1(r) , \quad A_1(t) = e^{\lambda t} A_1 .
		\end{split}
	\end{equation*}
	After algebraic manipulations, we get that the perturbation obeys to
	\begin{equation*}
		\begin{split}
			\lambda q_1(r) +\frac{d}{dr}q_1(r)=-\left(S(h_{\infty} , r)  +\lambda \right)  q_1 (r)   -J_s A_1   \frac{\partial S}{\partial h}( h_{\infty} , r)q_{\infty}(r) \hat \kappa (\lambda)  ,
		\end{split}
	\end{equation*}
	where we  have introduced $\hat \kappa$ the Laplace transform $\kappa$:
	\begin{equation*}
		\hat \kappa (\lambda) = \int_0^{\infty} \kappa (s)\exp (- \lambda s) \,ds,
	\end{equation*}
	and for the activity
	\begin{equation*}
		\begin{split}
			A_1\left(1-J_s\hat \kappa (\lambda) \int_{0}^{+\infty}  \frac{\partial S}{\partial h} (h_{\infty} , r)q_{\infty}(r) \,dr \right)  =\int_{0}^{+\infty}S( h_{\infty} ,r)q_1(r) \,dr  .
		\end{split}
	\end{equation*}
	Integrating this solution with the variation of constants method, we get 
	\begin{equation*}
		\begin{split}
			q_1(r) =  A_1 e^{- \int_0^{r}  S(h_{\infty} , s)  +\lambda  \,ds }  -J_s A_1 \hat \kappa (\lambda)  \int_0^{r}    \frac{\partial S}{\partial h}(h_{\infty},x)q_{\infty}(x) e^{- \int_x^{r}  S(h_{\infty}, s)  +\lambda  \,ds } \,dx,
		\end{split}
	\end{equation*}
	which implies 
	\begin{equation*}
		\begin{split}
			\int_{0}^{+\infty}S(h_{\infty}, r)q_1(r) \,dr = 
			-J_s A_1 \hat \kappa (\lambda)   \int_{0}^{+\infty}S(h_{\infty}, x) \int_0^{r}    \frac{\partial S}{\partial h}(h_{\infty} , x)q_{\infty}(x) e^{- \int_x^{r}  S(h_{\infty}, s)  +\lambda  \,ds } \,dx  \,dr \\
			+A_1  \int_{0}^{+\infty}S(h_{\infty} , r) e^{- \int_0^{r}  S(h_{\infty}, s)  +\lambda  \,ds } \,dr ,
		\end{split}
	\end{equation*}
	and we finally arrive on the equation
	
	\begin{equation*}
		\begin{split}
			1-J_s \hat \kappa (\lambda) \int_{0}^{+\infty}  \frac{\partial S}{\partial h} (h_{\infty} , r)q_{\infty}(r) \,dr +J_s  \hat \kappa (\lambda)   \int_{0}^{+\infty}S( h_{\infty}, r) \int_0^{r}   \frac{\partial S}{\partial h}( h_{\infty}, x)q_{\infty}(x) e^{- \int_x^{r}  S( h_{\infty} ,s)  +\lambda  \,ds } \,dx  \,dr \\
			-  \int_{0}^{+\infty}S( h_{\infty} ,r) e^{- \int_0^{r}  S( h_{\infty}, s)  +\lambda  \,ds } \,dr =0.
		\end{split}
	\end{equation*}
	We therefore write down the characteristic equation of the eigenvalues as
	\begin{equation*}
		\begin{split}
			\mathcal{C} (\lambda)= 1-J_s \hat \kappa (\lambda) \int_{0}^{+\infty}  \frac{\partial S}{\partial h} (h_{\infty} , r)q_{\infty}(r) \,dr  -  \int_{0}^{+\infty}S(h_{\infty} ,r) e^{- \int_0^{r}  S( h_{\infty}, s)  +\lambda  \,ds } \,dr & \\
			+J_s  \hat \kappa (\lambda)  \int_{0}^{+\infty}S( h_{\infty}, r) \int_0^{r}  \frac{\partial S}{\partial h} ( h_{\infty}, x)q_{\infty}(x) e^{- \int_x^{r}  S( h_{\infty} ,s)  +\lambda  \,ds } \,dx  \,dr &. 
		\end{split}
	\end{equation*}

	With the special choice
	\begin{equation*}
		\begin{split}
			S(h_{\infty},r)  = e^{h_{\infty}} H(r-T_{ref})  ,
		\end{split}
	\end{equation*}
	we can push further the computation, and after algebraic manipulations, we find:
	
	\begin{equation*}
		\begin{split}
			\mathcal{C} (\lambda)= 1-J_s \hat \kappa (\lambda)  A_{\infty}  -  \dfrac{e^{h_{\infty} -\lambda T_{ref}}}{\lambda +h_{\infty} }   +J_s \hat \kappa (\lambda) \dfrac{ A_{\infty}e^{h_{\infty}}}{\lambda +h_{\infty} }.
		\end{split}
	\end{equation*}

	The bifurcation line, which separates an oscillatory dynamic from an asynchronous regime, can be obtained numerically by solving 
	\begin{equation*}
		\begin{split}
			\mathcal{C} (i \omega)= 0.
		\end{split}
	\end{equation*}
	
	%======   APPENDIX C  ==========%
	\section{The Adjoint Equation}\label{A03}
	To compute the PRC, we first rewrite the synaptic filtering as a differential equation. Having 
	\begin{equation*}
		\begin{split}
			I_{s}(t)= J_s\kappa *A(t) , \quad \kappa(t) = \dfrac{e^{-t/ \tau_s}}{\tau_s},
		\end{split}
	\end{equation*}
	is equivalent as having:
	\begin{equation*}
		\begin{split}
			\tau_s \frac{d}{d t}  I_{s}(t)= -I_{s}(t)   +J_sA(t) .
		\end{split}
	\end{equation*}
	We then assume that there is a stable oscillatory solution $(q_{o},I_{s_o})$ of period $T$ for the mean-field equation. Considering a small perturbation around the stable solution, we write 
	\begin{equation*}
		\begin{split}
			q(t,r) = q_{o}(t,r)+ \varepsilon q_p(t,r) +\mathcal{O}( \varepsilon^2 ), \quad I_s(t) = I_{s_o}(t)+ \varepsilon I_{s_p}(t) +\mathcal{O}( \varepsilon^2 ).
		\end{split}
	\end{equation*}
	Plugging these expressions and only keeping the first order term, we get that the perturbation obeys to the following set of equations
	\begin{equation*}
		\begin{split}
			\frac{\partial}{\partial t}q_p(t,r) +\frac{\partial}{\partial r}q_p(t,r)=-S(h_o(t),r)q_p(t,r)  -  \frac{\partial S}{\partial h} (h_{o}(t),r)q_{o}(t,r)  I_{s_p}(t),
		\end{split}
	\end{equation*}
	where
	\begin{equation*}
		\begin{split}
			h_o(t) = I_{ext} + I_{s_o}(t) ,
		\end{split}
	\end{equation*}
	and for the activity
	\begin{equation*}
		\begin{split}
			A_p(t)=\int_{0}^{+\infty}S(h_{o}(t),r)q_p(t,r) \,dr +  I_{s_p}(t) \int_{0}^{+\infty}  \frac{\partial S}{\partial h} (h_{o}(t),r)q_{o}(t,r) \,dr ,
		\end{split}
	\end{equation*}
	the boundary condition follows as
	\begin{equation*}
		\begin{split}
			q_p(t,0)=A_p(t),
		\end{split}
	\end{equation*}
	with
	\begin{equation*}
		\begin{split}
			\tau_s \frac{d}{d t}  I_{s_p}(t)= -I_{s_p}(t)   +J_s A_p(t) .
		\end{split}
	\end{equation*}
	Now, we can define a bilinear form as
	\begin{equation*}
		\left\langle 
		\left(
		\begin{array}{c}
			q_1\\
			I_1\\
		\end{array}
		\right)  , 
		\left(
		\begin{array}{c}
			q_2\\
			I_2\\
		\end{array}
		\right);t
		\right\rangle 
		= \int_{0}^{+\infty}  q_1(t,r)q_2(t,r)     \,dr      + I_1(t)I_2(t)    .
	\end{equation*}
	The PRC  $\left( Z_q, Z_I \right)$   would be given by the following property
	\begin{equation*}
		\begin{split}
			\frac{d}{d t}  \left\langle 
			\left(
			\begin{array}{c}
				Z_q\\
				Z_{I_s}\\
			\end{array}
			\right)  , 
			\left(
			\begin{array}{c}
				q_p\\
				I_p\\
			\end{array}
			\right);t
			\right\rangle =0  .
		\end{split}
	\end{equation*}
	Developing the first term we get that
	\begin{equation*}
		\begin{split}
			\frac{d}{d t}     \int_{0}^{+\infty}  Z_q(t,r)q_p(t,r)     \,dr  =    \int_{0}^{+\infty} q_p(t,r)\frac{\partial }{\partial t}Z_q (t,r)    + Z_q(t,r)  \frac{\partial }{\partial t}q_p (t,r)    \,dr   ,  
		\end{split}
	\end{equation*}
	and plugging the expression of $\frac{\partial }{\partial t}q_p (t,r)$ inside the equation, we obtain
	\begin{equation*}
		\begin{split}
			\frac{d}{d t}     \int_{0}^{+\infty}  Z_q(t,r)q_p(t,r)     \,dr  =    \int_{0}^{+\infty} Z_q(t,r)  \left( - \frac{\partial}{\partial r}q_p(t,r)-S(h_{o}(t),r)q_p(t,r)  -  \frac{\partial S}{\partial h} (h_{o}(t),r)q_{o}(t,r)  I_{s_p}(t) \right)  \,dr  \\
			+  \int_{0}^{+\infty} q_p(t,r)\frac{\partial }{\partial t}Z_q (t,r)      \,dr  ,
		\end{split}
	\end{equation*}
	developing the terms lead to 
	\begin{equation*}
		\begin{split}
			\frac{d}{d t}     \int_{0}^{+\infty}  Z_q(t,r)q_p(t,r)     \,dr   =     -  \int_{0}^{+\infty} Z_q(t,r)  \frac{\partial}{\partial r}q_p(t,r) \,dr -    \int_{0}^{+\infty} Z_q(t,r)  S(h_{o}(t),r)q_p(t,r)  \,dr \\
			-   I_{s_p}(t)  \int_{0}^{+\infty} Z_q(t,r)  \frac{\partial S}{\partial h} (h_{o}(t),r)q_{o}(t,r)   \,dr  +  \int_{0}^{+\infty} q_p(t,r)\frac{\partial }{\partial t}Z_q (t,r)      \,dr . \\
		\end{split}
	\end{equation*}
	Applying an integration by parts we get
	\begin{equation*}
		\begin{split}
			\int_{0}^{+\infty}  Z_q(t,r)  \frac{\partial}{\partial r}q_p(t,r)  \,dr  = &  \left[ Z_q(t,r)  q_p(t,r)  \right]_{0}^{+\infty}  -  \int_{0}^{+\infty}  \frac{\partial}{\partial r} Z_q(t,r)  q_p(t,r)  \,dr \\ 
			= &   -   Z_q(t,0)  q_p(t,0)     -  \int_{0}^{+\infty}  \frac{\partial}{\partial r} Z_q(t,r)  q_p(t,r)  \,dr \\ 
			= &   -   Z_q(t,0)  A_p(t)    -  \int_{0}^{+\infty}  \frac{\partial}{\partial r} Z_q(t,r)  q_p(t,r)  \,dr  .
		\end{split}
	\end{equation*}
	Therefore we have
	\begin{equation*}
		\begin{split}
			\frac{d}{d t}     \int_{0}^{+\infty}  Z_q(t,r)q_p(t,r)     \,dr   
			=       Z_q(t,0)  A_p(t)    +  \int_{0}^{+\infty}  \frac{\partial}{\partial r} Z_q(t,r)  q_p(t,r)  \,dr -    \int_{0}^{+\infty} Z_q(t,r)  S(h_{o}(t),r)q_p(t,r)  \,dr  \\
			-   I_{s_p}(t)  \int_{0}^{+\infty} Z_q(t,r)  \frac{\partial S}{\partial h} (h_{o}(t),r)q_{o}(t,r)   \,dr  +  \int_{0}^{+\infty} q_p(t,r)\frac{\partial }{\partial t}Z_q (t,r)      \,dr  ,
		\end{split}
	\end{equation*}
	which is equivalent to
	\begin{equation*}
		\begin{split}
			\frac{d}{d t}     \int_{0}^{+\infty}  Z_q(t,r)q_p(t,r)     \,dr   
			=        \int_{0}^{+\infty} \left( \frac{\partial }{\partial t}Z_q (t,r) +   \frac{\partial}{\partial r} Z_q(t,r) -S(h_{o}(t),r)Z_q(t,r) \right)   q_p(t,r)  \,dr \\
			+ Z_q(t,0)  A_p(t)    -   I_{s_p}(t)  \int_{0}^{+\infty} Z_q(t,r)  \frac{\partial S}{\partial h} (h_{o}(t),r)q_{o}(t,r)   \,dr    ,
		\end{split}
	\end{equation*}
	We now develop the second term 
	\begin{equation*}
		\begin{split}
			\frac{d}{d t}   \left[   Z_{I_s}(t)I_{s_p}(t) \right] =I_{s_p}(t)\frac{d }{d t}  Z_{I_s}(t) +  Z_{I_s}(t) \frac{d}{d t}I_{s_p} (t)  ,
		\end{split}
	\end{equation*}
	and recalling the fact that
	\begin{equation*}
		\begin{split}
			\tau_s \frac{d}{d t}  I_{s_p}(t)= -I_{s_p}(t)   +J_s A_p(t) ,
		\end{split}
	\end{equation*}
	we obtain
	\begin{equation*}
		\begin{split}
			\frac{d}{d t}   \left[   Z_{I_s}(t)I_{s_p}(t) \right] =I_{s_p}(t)\frac{d }{d t}  Z_{I_s}(t) - \dfrac{1}{\tau_s } Z_{I_s}(t) I_{s_p}(t)  +\dfrac{J_s}{\tau_s }  Z_{I_s}(t)A_p(t)   .
		\end{split}
	\end{equation*}
	Now, putting everything together
	\begin{equation*}
		\begin{split}
			\frac{d}{d t}  \left\langle 
			\left(
			\begin{array}{c}
				Z_q\\
				Z_{I_s}\\
			\end{array}
			\right)  , 
			\left(
			\begin{array}{c}
				q_p\\
				I_{s_p}\\
			\end{array}
			\right);t
			\right\rangle =\frac{d}{d t}  \int_{0}^{+\infty}  Z_q(t,r)q_p(t,r)     \,dr      +  \frac{d}{d t} \left[  Z_{I_s}(t)I_{s_p}(t) \right]  ,
		\end{split}
	\end{equation*}
	which gives
	\begin{equation*}
		\begin{split}
			\frac{d}{d t}  \left\langle 
			\left(
			\begin{array}{c}
				Z_q\\
				Z_{I_s}\\
			\end{array}
			\right)  , 
			\left(
			\begin{array}{c}
				q_p\\
				I_{s_p}\\
			\end{array}
			\right);t
			\right\rangle =\int_{0}^{+\infty} \left( \frac{\partial }{\partial t}Z_q (t,r) +   \frac{\partial}{\partial r} Z_q(t,r) -S(h_{o}(t),r)Z_q(t,r) \right)   q_p(t,r)  \,dr \\
			+ Z_q(t,0)  A_p(t)    -   I_{s_p}(t)  \int_{0}^{+\infty} Z_q(t,r)  \frac{\partial S}{\partial h} (h_{o}(t),r)q_{o}(t,r)   \,dr \\
			+I_{s_p}(t)\frac{d }{d t}  Z_{I_s}(t) - \dfrac{1}{\tau_s } Z_{I_s}(t) I_{s_p}(t)+ \dfrac{J_s}{\tau_s}  Z_{I_s}(t ) A_p(t)   .
		\end{split}
	\end{equation*}
	We now use the fact that
	\begin{equation*}
		\begin{split}
			A_p(t)=\int_{0}^{+\infty}S(h_{o}(t),r)q_p(t,r) \,dr +  I_{s_p}(t) \int_{0}^{+\infty}  \frac{\partial S}{\partial h} (I_{o}(t),r)q_{o}(t,r) \,dr ,
		\end{split}
	\end{equation*}
	we obtain
	\begin{equation*}
		\begin{split}
			\int_{0}^{+\infty} \left( \frac{\partial }{\partial t}Z_q (t,r) +   \frac{\partial}{\partial r} Z_q(t,r) -S(h_{o}(t),r)  \left(  Z_q(t,r) - Z_q(t,0)   -\dfrac{J_s}{\tau_s }   Z_{I_s}(t)  \right)    \right)   q_p(t,r)  \,dr \\
			+     I_{p_s}(t)   \left(  \frac{d }{d t}  Z_{I_s}(t) -  \dfrac{1}{\tau_s } Z_{I_s}(t) - \int_{0}^{+\infty} \left( Z_q(t,r)  -Z_q(t,0)   -\dfrac{J_s}{\tau_s }   Z_{I_s}(t) \right)   \frac{\partial S}{\partial h} (h_{o}(t),r)q_{o}(t,r)   \,dr    \right) =0  .
		\end{split}
	\end{equation*}
	Since this is true for every perturbation, the PRC must solve
	\begin{equation}\label{iPRC1sup}
		\begin{split}
			-\frac{\partial }{\partial t}Z_q (t,r)  - \frac{\partial}{\partial r}Z_q(t,r) = - S(h_{o}(t),r)  \left[ Z_q (t,r)  - Z_q (t,0) -   \dfrac{J_s}{\tau_s} Z_{I_s}(t)  \right] ,
		\end{split}
	\end{equation}
	and
	\begin{equation}\label{iPRC2sup}
		\begin{split}
			-   \frac{d }{d t} Z_{I_s} (t)   =-  \dfrac{1}{\tau_s} Z_{I_s}(t)     -   \int_{0}^{+\infty}\left[ Z_q (t,r)  - Z_q (t,0) -   \dfrac{J_s}{\tau_s} Z_{I_s}(t)  \right]  \frac{\partial S}{\partial h} (h_{o}(t),r)q_{o}(t,r)   \,dr.
		\end{split}
	\end{equation}\\

	%======   APPENDIX E  ==========%
	\section{Normalization condition} \label{A04}
	The adjoint equation being linear, its solution is unique under a normalization condition.
	In what follows we check that
	\begin{equation*}
		\begin{split}
			\frac{d}{d t}  \left\langle 
			\left(
			\begin{array}{c}
				Z_q\\
				Z_{I_s}\\
			\end{array}
			\right)  , 
			\left(
			\begin{array}{c}
				\frac{\partial }{\partial t}q_o\\
				\frac{d }{d t} I_{s_o} \\
			\end{array}
			\right);t
			\right\rangle =0  .
		\end{split}
	\end{equation*}
	The computations that fallow give rise to long mathematical expressions. We thus drop the function variables. After algebraic manipulations, we find that the above condition is equivalent to
	\begin{equation*}
		\begin{split}
			\int_{0}^{+\infty}  \frac{\partial }{\partial t} Z_q \frac{\partial }{\partial t}q_o     \,dr      +  \frac{d}{d t}Z_{I_s}  \frac{d }{d t} I_{s_o}    +
			\int_{0}^{+\infty}  \frac{\partial }{\partial t} Z_q \frac{\partial }{\partial t} \left( - \frac{\partial }{\partial r}q_o -  S_oq_o  \right)   \,dr      +  \frac{d}{d t}Z_{I_s}  \frac{\partial }{\partial t} \left(  -  \dfrac{1}{\tau_s} {I_{s_o}} + \dfrac{J_s}{\tau_s} A_o      \right) =0.
		\end{split}
	\end{equation*}
	where we have introduced the new notations:
	\begin{equation*}
		A_o:= \int_{0}^{+\infty} S_o q_o     \,dr      \quad S_o:= S(h_o(t),r) .
	\end{equation*}
	Now developing, we get
	\begin{equation*}
		\begin{split}
			& \int_{0}^{+\infty}  \frac{\partial }{\partial t} Z_q \frac{\partial }{\partial t} \left( - \frac{\partial }{\partial r}q_o -  S_o q_o  \right)   \,dr      \\
			=& \int_{0}^{+\infty}  \frac{\partial }{\partial t} Z_q \left( - \frac{\partial }{\partial r }  \frac{\partial }{\partial t} q_o -  S_o   \frac{\partial }{\partial t}q_o -    \frac{\partial S_o}{\partial h}  q_o \frac{d }{d t}I_{s_o} \right)   \,dr  \\
			=& \int_{0}^{+\infty}  \frac{\partial }{\partial t} \frac{\partial }{\partial r }   Z_q \frac{\partial }{\partial t} q_o - Z_q S_o   \frac{\partial }{\partial t}q_o - Z_q   \frac{\partial S_o}{\partial h}  q_o \frac{d }{d t}I_{s_o}   \,dr -\left[ Z_q \frac{\partial }{\partial t} q_o    \right]_0^{+\infty}   \\
			=& \int_{0}^{+\infty}  \frac{\partial }{\partial t} \frac{\partial }{\partial r }   Z_q \frac{\partial }{\partial t} q_o - Z_q S_o   \frac{\partial }{\partial t}q_o - Z_q   \frac{\partial S_o}{\partial h}  q_o \frac{d }{d t}I_{s_o}   \,dr + Z_q (t,0) \frac{d }{d t}A_{o}. 
		\end{split}
	\end{equation*}
	We now use the fact that
	\begin{equation*}
		\begin{split}
			\frac{d }{d t}   A_{o}= \int_{0}^{+\infty}  S_o   \frac{\partial }{\partial t}q_o \,dr +  \int_{0}^{+\infty}    \frac{\partial S_o}{\partial h}  q_o \frac{d }{d t}I_{s_o}   \,dr .
		\end{split}
	\end{equation*}
	Using this expression, we get that
	\begin{equation*}
		\begin{split}
			\int_{0}^{+\infty}  \frac{\partial }{\partial t}  Z_q \frac{\partial }{\partial t}  \left( - \frac{\partial }{\partial r}q_o -  S_oq_o  \right) &  \,dr      +  \frac{d}{d t}Z_{I_s}  \frac{d}{dt} \left(  -  \dfrac{1}{\tau_s} {I_{s_o}} + \dfrac{J_s}{\tau_s} A_o      \right) \\
			=      \int_{0}^{+\infty}  \frac{\partial }{\partial t}   &q_o   \left(   \frac{\partial }{\partial r } Z_q   -S_o  Z_q   +Z_q(t,0)S_o  +\dfrac{J_s}{\tau_s} S_o Z_{I_s}  \right)     \,dr  \\
			&+\frac{d}{d t} I_o \left( Z_{I_s} / \tau_s   - \int_{0}^{+\infty} \frac{\partial S_o}{\partial h}  q_o \left[    Z_q   -Z_q(t,0)  -\dfrac{J_s}{\tau_s}Z_{I_s} \right]  \,dr  \right).
		\end{split}
	\end{equation*}
	Putting everything together, we arrive to
	\begin{equation*}
		\begin{split}
			\frac{d }{d t}  \left[ \int_{0}^{+\infty}   Z_q \frac{\partial }{\partial t}q_o      \,dr   \right. &  \left.+  Z_{I_s}  \frac{d }{d t} I_{s_o} \right]  \\
			=      \int_{0}^{+\infty}  \frac{\partial }{\partial t} &  q_o   \left(    \frac{\partial }{\partial t } Z_q +\frac{\partial }{\partial r } Z_q   -S_o  Z_q   +Z_q(t,0)S_o  +\dfrac{J_s}{\tau_s} S_o Z_{I_s}  \right)     \,dr \\
			& +\frac{d}{d t} I_{s_o} \left( \frac{d}{d t} Z_{I_s}   -Z_{I_s} / \tau_s   - \int_{0}^{+\infty} \frac{\partial S_o}{\partial h}  q_o \left[    Z_q   -Z_q(t,0)  -\dfrac{J_s}{\tau_s}Z_{I_s} \right]  \,dr  \right).
		\end{split}   
	\end{equation*}
	We now remind that the adjoint system is given by
	\begin{equation*}
		\begin{split}
			-   \frac{\partial }{\partial t } Z_q -\frac{\partial }{\partial r } Z_q   =-S_o  Z_q   +Z_q(t,0)S_o  +\dfrac{J_s}{\tau_s} S_o Z_{I_s}      ,
		\end{split}
	\end{equation*}
	and
	\begin{equation*}
		\begin{split}
			-\frac{d}{d t} Z_{I_s}   -Z_{I_s} / \tau_s   - \int_{0}^{+\infty} \frac{\partial S_o}{\partial h}  q_o \left[    Z_q   -Z_q(t,0)  -\dfrac{J_s}{\tau_s}Z_{I_s} \right]  \,dr ,
		\end{split}
	\end{equation*}
	we therefore arrive to
	\begin{equation*}
		\frac{d}{d t}    \left[ \int_{0}^{+\infty}  Z_q \frac{\partial }{\partial t}q_o     \,dr      +  Z_{I_s}  \frac{d }{d t} I_{s_o}  \right]  =  0.
	\end{equation*}
	The iPRC will be the unique solution satisfying the normalization condition:
	\begin{equation*}
		\int_{0}^{+\infty}  Z_q \frac{\partial }{\partial t}q_o     \,dr      +  Z_{I_s}  \frac{d }{d t} I_{s_o}    =  \dfrac{2\pi}{T},
	\end{equation*}
	where $T$ is nothing period of the oscillation. \\

	%======   APPENDIX E  ==========%
	\section{Numerical procedure} \label{A05}
	The mean-field equation (\ref{ASsup}) can be readily integrated. We denote
	\begin{align*}
		r_j = j \Delta t, \quad \forall j > 0, \quad   t_n = n \Delta t, \quad \forall n > 0,
	\end{align*}
	the discretization space/time variables, and
	\begin{align*}
		q^n_j := q\left(  t^n ,  r_j  \right), \quad S^n_j  :=S \left( h^n ,  r_j  \right), \quad  h^n:=h(t^n), \quad  I^n_{ext}:=I_{ext}(t^n) \quad  I^n_{s}:=I_{s}(t^n),
	\end{align*}
	the corresponding solution at the discretized points.
	
	Considering the initial state to be given, the mean-field equation (\ref{AS}) can be numerically solved along the characteristic curves. On the characteristics, the dynamics reduce to a
	nonlinear differential equation that can be integrated with the following first order numerical scheme:
	\begin{equation} \label{NS}
		\left \{
		\begin{split}
			&q^{n+1}_{j+1}  = q^n_j  -  \Delta t S^n_j q^n_j     \\
			&I_s^{n+1}=I_s^{n} +\Delta t  \left( -I_s^{n}/ \tau_s+ J_sA^{n}/ \tau_s \right)\\
			&q^{n+1}_{1} = A^{n+1}\\
			&A^{n+1}=\Delta t  \sum_{k \ge 1}  S^{n+1}_j q^{n+1}_j   \\
			&h^{n+1}= I_{ext}^{n+1}+   I_s^{n+1} .\\
		\end{split}
		\right .
	\end{equation}
	The proposed numerical scheme (\ref{NS}) is thus well defined and produces results in excellent agreement with
	simulations of the full network.
	
	Using procedure \eqref{NS} we find solutions of period $T = M\Delta t$ for the mean mean-field equation \eqref{ASsup} which we denote as $\bar{q}(t,r)$ and $\bar{I}_{s}(t,r)$. Next, we use the solutions $\bar{q}(t,r)$ and $\bar{I}_{s}(t,r)$ for solving the adjoint system (\ref{iPRC1sup})-(\ref{iPRC2sup}).

	Since the solution of the adjoint equation has an opposite stability with respect to the mean-field, we must integrate it backwards in time. We denote
	\begin{align*}
		Z^n_{q_j} := Z_q\left(  t^n ,  r_j  \right), \quad   Z_{I_s}^n:=Z_{I_s}(t^n), \quad \bar{S}^n_j  :=S \left( \bar{h}^n ,  r_j  \right) \quad \partial \bar{S}^n_j:=\dfrac{\partial S}{\partial h} \left( \bar{h}^n ,  r_j  \right), \quad  \bar{h}^n = I_{ext}^{n}+   \bar{I}_s^{n}.
	\end{align*}
	Considering the end state to be given, the adjoint system (\ref{iPRC1sup})-(\ref{iPRC2sup}) can be once again numerically solved along the characteristic curves. On the characteristics, the dynamics of the adjoint system (\ref{iPRC1sup})-(\ref{iPRC2sup}) reduce to a
	linear differential equation that can be integrated with the following backward first order numerical scheme:
	\begin{equation} \label{ANS}
		\left \{
		\begin{split}
			&Z^{n-1}_{q_{j-1}} = Z^n_{q_j} -  \Delta t   \bar{S}^n_j \left[ Z^n_{q_j} - Z^n_{q_1} - J_sZ_{I_s}^n/\tau_s     \right]  \\
			&Z^{n-1}_{q_{l}} = Z^{n-1}_{q_{l-1}} \qquad \qquad \qquad \qquad \qquad \qquad \qquad \qquad \text{for} \enskip l = \max(j) \\
			&Z_{I_s}^{n-1}  = Z_{I_s}^n  - \Delta t \Big(  Z_{I_s}^n / \tau_s + \sum_{k \ge 1}    \partial \bar{S}^n_k   \left[    Z^n_{q_k} -Z^n_{q_1} -  J_sZ_{I_s}^n/\tau_s   \right]  \bar{q}^n_k \Delta t \Big) .\\
		\end{split}
		\right .
	\end{equation}
	
	The proposed numerical scheme \eqref{ANS} is once again well defined and produces $T$ periodic solutions $\bar{Z}_q(t,r)$ and $\bar{Z}_{I_s}(t)$ matching
	the PRC obtained by the direct perturbation method (see the main text). Next, we remark some numerical recipes which enhance the stability (and thus the convergence) of the procedure in \eqref{ANS}. First, we iterate the scheme \eqref{ANS} over the periodic solutions $\bar{q}(t,r)$ and $\bar{I}_{s}(t,r)$ (recall $\bar{q}^{n+M}_k = \bar{q}^{n}_k$). We also recommend computing the integral in \eqref{iPRC2sup} (that is, the sum for $Z_{I_s}^{n-1}$ in \eqref{ANS}) by using precise integration routines such as the trapezoidal rule or the Simpson's method. Finally, since the procedure in \eqref{ANS} is based on backwards integration, it does not provide the value of $Z_q(t_n, r_j)$ at $r = \max(r_j)$. This value can be obtained by simple extrapolation (as we propose in \eqref{ANS}) or by using accurate extrapolation routines taking into account a larger set of values of $Z_q(t_n, r_j)$. We remark that although the smaller the $\Delta t$ value, the higher the accuracy of solutions, the usage of the above mentioned recipes generates very precise results for time steps around $\Delta t =0.005$.\\

	\section{Phase-resetting curve for an excitatory-inhibitory network}
	
	In the thermodynamic limit the network description of a pair of excitatory-inhibitory populations reduces to a set of coupled partial differential equations. 
	Denoting $q_e(t,r)$ the probability density for a excitatory neuron to have at time $t$ an age $r$, and  $q_i(t,r)$ for the inhibitory population, the evolution of the density profiles evolve according to the continuity equations:
	\begin{equation*}
		\begin{split}
			\frac{\partial}{\partial t}q_e +\frac{\partial}{\partial r}q_e=-S_e(h_e(t),r)q_e,
		\end{split}
	\end{equation*}
	and
	\begin{equation*}
		\begin{split}
			\frac{\partial}{\partial t}q_i +\frac{\partial}{\partial r}q_i=-S_i(h_i(t),r)q_i.
		\end{split}
	\end{equation*}
	The boundary conditions are given by
	\begin{equation*}
		q_e(t,0)=A_e(t)=\int_{0}^{+\infty}S_e(h_e(t),r)q_e(t,r) \,dr , 
	\end{equation*}
	and 
	\begin{equation*}
		q_i(t,0)=A_i(t)=\int_{0}^{+\infty}S_i(h_i(t),r)q_i(t,r) \,dr .
	\end{equation*}
	The total input current is still given by 
	\begin{equation*}
		h_e = I_{ext} ^e    + I_{s_e}  ,  \quad h_i = I_{ext} ^i    + I_{s_i}  ,
	\end{equation*}
	the synaptic current $I_{s}(t)$ is computed as
	\begin{equation*}
		\begin{split}
			I_{s_e}=   J_{ee} \kappa * A_e-J_{ei} \kappa * A_i  , \quad  I_{s_i}=   J_{ie} \kappa * A_e-J_{ii} \kappa * A_i  .
		\end{split}
	\end{equation*}
	We can now define the corresponding bi-linear form:
	\begin{equation*}
		\left\langle 
		\left(
		\begin{array}{c}
			q_{e_1}\\
			I_{e_1}\\
			q_{i_1}\\
			I_{i_1}\\
		\end{array}
		\right)  , 
		\left(
		\begin{array}{c}
			q_{e_2}\\
			I_{e_2}\\
			q_{i_2}\\
			I_{i_2}\\
		\end{array}
		\right);t
		\right\rangle 
		= \int_{0}^{+\infty}  q_{e_1}q_{e_2}     \,dr      + I_{e_1}I_{e_2} + \int_{0}^{+\infty}  q_{i_1}q_{i_2}     \,dr      + I_{i_1}I_{i_2}    .
	\end{equation*}
	Assuming to be known the periodic solution, $(q_{e_o},I_{s_{e_o}})$ and $(q_{i_o},I_{s_{i_o}})$, computations similar to what is presented within the adjoint section, we find that the PRC must solves:
	\begin{equation*}
		\begin{split}
			-\frac{\partial }{\partial t}Z_{q_e}   - \frac{\partial}{\partial r}Z_{q_e}  = - S_e(h_{e_o}(t),r)  \left[ Z_{q_e}    - Z_{q_e}  (t,0) -   \dfrac{J_{ee}}{\tau_s} Z_{I_{s_e}} +   \dfrac{J_{ei}}{\tau_s} Z_{I_{s_i}}  \right] ,
		\end{split}
	\end{equation*}
	and
	\begin{equation*}
		\begin{split}
			-   \frac{d }{d t} Z_{I_{s_e}}    =-  \dfrac{1}{\tau_s} Z_{I_{s_e}}    & -   \int_{0}^{+\infty}\left[ Z_{q_e}   - Z_{q_e} (t,0) -   \dfrac{J_{ee}}{\tau_s} Z_{I_{s_e}}  \right]  \frac{\partial S_e}{\partial h_e} (h_{e_o}(t),r)q_{e_o}   \,dr\\
			&   -   \int_{0}^{+\infty}\left[ Z_{q_i}   - Z_{q_i} (t,0) +   \dfrac{J_{ei}}{\tau_s} Z_{I_{s_i}}   \right]  \frac{\partial S_i}{\partial h_i} (h_{i_o}(t),r)q_{i_o}   \,dr.
		\end{split}
	\end{equation*}
	similarly
	\begin{equation*}
		\begin{split}
			-\frac{\partial }{\partial t}Z_{q_i}   - \frac{\partial}{\partial r}Z_{q_i}  = - S_i(h_{i_o}(t),r)  \left[ Z_{q_i}    - Z_{q_i}  (t,0) -   \dfrac{J_{ie}}{\tau_s} Z_{I_{s_e}} +   \dfrac{J_{ii}}{\tau_s} Z_{I_{s_i}}  \right] ,
		\end{split}
	\end{equation*}
	and
	\begin{equation*}
		\begin{split}
			-   \frac{d }{d t} Z_{I_{s_i}}    =-  \dfrac{1}{\tau_s} Z_{I_{s_i}}    & -   \int_{0}^{+\infty}\left[ Z_{q_e}   - Z_{q_e} (t,0) -   \dfrac{J_{ie}}{\tau_s} Z_{I_{s_e}}  \right]  \frac{\partial S_e}{\partial h_e} (h_{e_o}(t),r)q_{e_o}   \,dr\\
			&   -   \int_{0}^{+\infty}\left[ Z_{q_i}   - Z_{q_i} (t,0) +   \dfrac{J_{ii}}{\tau_s} Z_{I_{s_i}}  \right]  \frac{\partial S_i}{\partial h_i} (h_{i_o}(t),r)q_{i_o}   \,dr.
		\end{split}
	\end{equation*}
	Incoming perturbation should get through the synapse, $Z_{I_s}$ should be interpreted as the  iPRC of the macroscopic oscillation. Two PRCs can therefore be defined $Z_{I_{s_e}}$  and $Z_{I_{s_i}}$ at the same time. 
	The PRC defined by $Z_{I_{s_e}}$ corresponds to excitatory input arriving upon the E-cells, while  $Z_{I_{s_i}}$ corresponds to excitatory input arriving upon the I-cells. 
	
	The normalisation condition is now given by:
	\begin{equation}\label{normalisation}
		\int_{0}^{+\infty}  Z_{q_e} \frac{\partial }{\partial t}q_{o_e}     \,dr      +  Z_{I_{s_e}} \frac{d }{d t} I_{s_{o_e}}   +  \int_{0}^{+\infty}  Z_{q_i} \frac{\partial }{\partial t}q_{o_i}     \,dr      +  Z_{I_{s_i}} \frac{d }{d t} I_{s_{o_i}}  =  \dfrac{2\pi }{T},
	\end{equation}
	with again $T$ the oscillation period.
	
	\noindent \textbf{Acknowledgements:} The research leading to these results has received funding from the Basic Research Program at the National Research University Higher School of Economics and the ANR Project ERMUNDY (Grant No ANR-18-CE37-0014). \\

	\bibliographystyle{abbrv}
	\bibliography{PRC}

\end{document}